\documentclass{article}
\usepackage{amsmath}
\usepackage{amsfonts}
\usepackage{graphicx}
\usepackage{bm}
\usepackage{mathtools}
\usepackage[margin=1in]{geometry}
\usepackage{authblk}
\usepackage{pdfpages}

\mathtoolsset{showonlyrefs}

\usepackage[authoryear]{natbib}
\title{Flexibly Modeling Shocks to Demographic and Health Indicators with Bayesian Shrinkage Priors}
\author[1]{Herbert Susmann}
\author[2]{Leontine Alkema}
\affil[1]{\small Division of Biostatistics, Department of Population
  Health, New York University Grossman School of Medicine}
\affil[2]{\small Department of Biostatistics \& Epidemiology, University of Massachusetts Amherst}
\date{}

\begin{document}

\maketitle

\begin{abstract}
Demographic and health indicators may exhibit short or large short-term shocks; for example, armed conflicts, epidemics, or famines may cause shocks in period measures of life expectancy. Statistical models for estimating historical trends and generating future projections of these indicators for a large number of populations may be biased or not well probabilistically calibrated if they do not account for the presence of shocks. We propose a flexible method for modeling shocks when producing estimates and projections for multiple populations. The proposed approach makes no assumptions about the shape or duration of a shock, and requires no prior knowledge of when shocks may have occurred. Our approach is based on the modeling of shocks in level of the indicator of interest. We use Bayesian shrinkage priors such that shock terms are shrunk to zero unless the data suggest otherwise. The method is demonstrated in a model for male period life expectancy at birth. We use as a starting point an existing projection model and expand it by including the shock terms, modeled by the Bayesian shrinkage priors. Out-of-sample validation exercises find that including shocks in the model results in sharper uncertainty intervals without sacrificing empirical coverage or prediction error.
\end{abstract}


\section*{Introduction}
International organizations monitor demographic and health indicators as a way to track progress towards meeting international goals and to identify populations where improvement is needed. Short- and long-term projections of indicators may be of interest to inform policy evaluation and planning. For example, male period life expectancy at birth for a given time period, referred to here as $e_0$, is defined as the life expectancy of a hypothetical male cohort if it were subject to the age-specific mortality rates of that period for its entire life \citep{guillot2011periodlife}. Estimation and projection of country-specific male period life expectancy at birth is of fundamental interest in itself as a key demographic indicator, and as an input to population projections \citep{raftery2014popprojections}. 

Statistical models are typically used to generate such projections, with the goal of making probabilistic statements that accurately characterize uncertainty. In the case of male life expectancy, projections are produced by the United Nations Population Division (UNPD) using the probabilistic projection model developed by \cite{raftery2013life}. In this model, the evolution over time in life expectancy is assumed to follow a transition from low to high values. The expected rate of change is allowed to vary with its level. A stochastic smoothing component is added to capture temporary stalls or accelerations.

When developing a statistical model to  estimate and project a demographic indicator of interest, the occurrence of extremely large and rapid changes needs to be considered. We refer to any large and rapid change in the level of an indicator as a \textit{shock}, where the threshold of what constitutes a shock will depend substantively on the indicator of interest. For example, life expectancy may decrease drastically during wars, epidemics, or famines. The UNPD's life expectancy projection model does not consider shocks explicitly. Fluctuations, including possibly extreme ones, are captured through stochastic smoothing terms that apply to all periods. Implicitly, projections produced by such methods include uncertainty due to shocks. 

Statistically, it is important to address shocks because their presence violates the assumptions of any model that presumes an indicator follows smooth temporal trends. Such a mismatch between the truth and model assumptions may manifest in biased projections and uncertainty intervals that are not well calibrated. In addition, modeling shocks explicitly is important if interest is in projections of an indicator conditional on shocks not occurring in the future. For example, instead of projections of male period life expectancy, the target of interest may be projections of \textit{shock-free} male period life expectancy, defined as what the period life expectancy would be in the future assuming the absence of mortality shocks. 

We propose a flexible method for modeling shocks when producing
projections for multiple populations, and apply the approach to project life expectancy. In our approach, no knowledge is needed about the origins of the shocks or when they may have occurred. Rather, we take a statistical approach based on Bayesian shrinkage priors in which we assume that, while a shock is possible at any time point during the period of observations, the vast majority of time points will not have a shock. When the data strongly support the presence of a shock, it will be captured by the model estimates; otherwise, the estimated shock term will be approximately zero. We show that life expectancy projections from a model with shocks improve upon those constructed by the same model without shock terms. 

The rest of the article is organized as follows. We first introduce the life expectancy data used in this paper. Next, we introduce the life expectancy projection method and describe how Bayesian shrinkage priors can be used within the framework to estimate shocks. We then present empirical results and compare results between models with and without shock terms. The discussion includes our thoughts on how this approach is generalizable to other settings.

\section*{Male Period Life Expectancy at Birth}
Period life expectancy at birth for a given time period is defined as the life expectancy of a hypothetical cohort if it were subject to the age-specific mortality rates of that period for its entire life \citep{guillot2011periodlife}. Estimation and projection of country-specific period life expectancy at birth of fundamental interest in itself as a key demographic indicator, and as an input to population projections \citep{raftery2014popprojections}. We focus on male period life expectancy at birth, which we denote $e_0$ and for brevity may refer to as simply ``life expectancy".

The United Nations Population Division (UNPD) regularly produces estimates and projections of $e_0$ in countries. A recent set is found in UNPD World Population Prospects (WPP) 2022 revision \citep{wpp2022methodology}, with estimates and projections publicly available in the \texttt{wpp2022} R package \citep{Rwpp2022}. For our study, we use the UNPD's 5-year period estimates from 1950-1955 to 2015-2020 for 208 countries with population greater than 1,000,000 in the year 2020. % We analyze the same subset of countries as \cite{raftery2013life}. 
Following the approach by \cite{raftery2013life}, this set of countries excludes 38 countries subject to an HIV/AIDS epidemic because modeling life expectancy in these contexts requires additional considerations (see \cite{godwin2017hiv}) not relevant to our present focus on short-term shocks. We do not include the period 2020-2025 in our set of estimates because this period was mostly in the future when the estimates were produced in 2022, and hence informed by very limited data. The use of this cut off implies that shocks due to the Covid-19 pandemic are not considered (see discussion). 

\begin{figure}[!htbp]
    \includegraphics[width=\columnwidth]{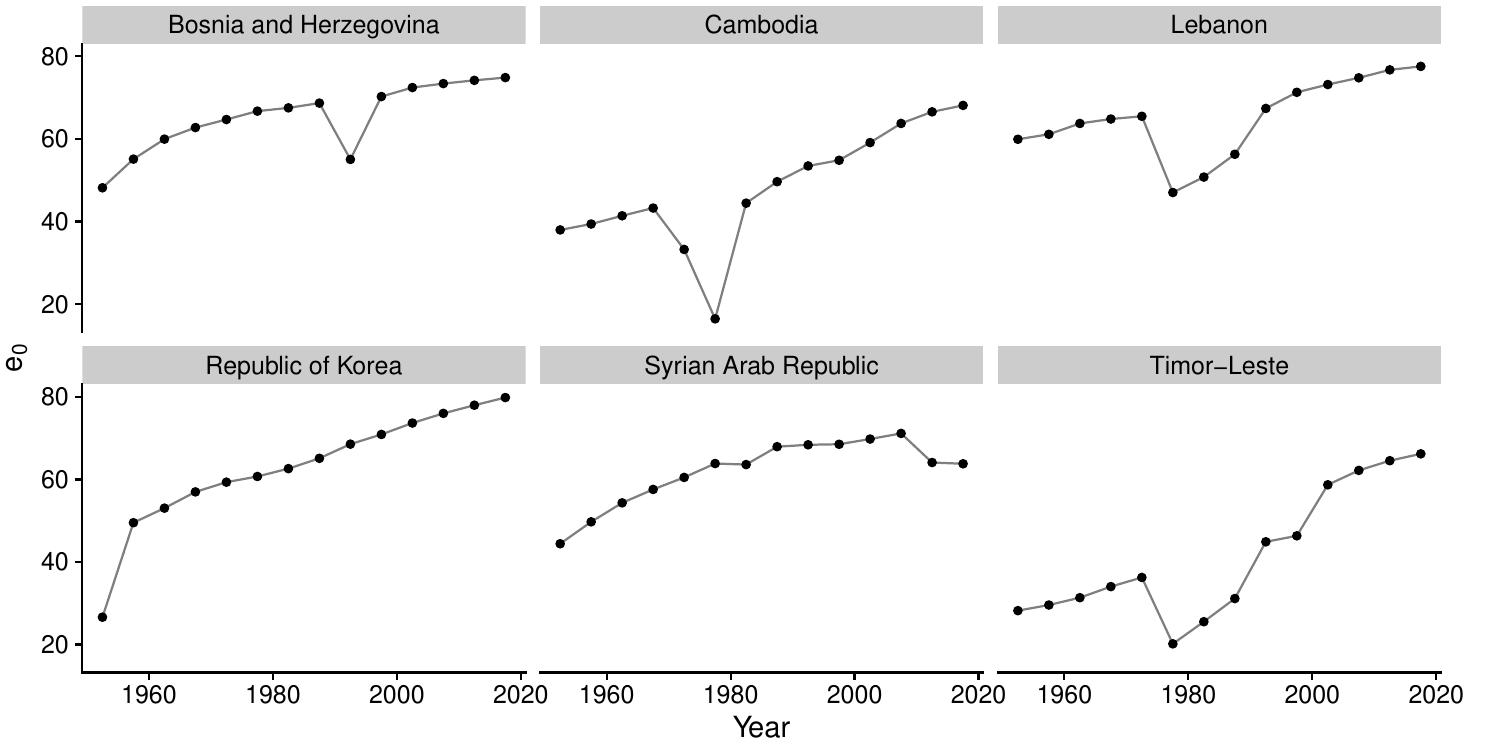}
    \caption{\label{fig:life-examples} Estimated male period life expectancy at birth ($e_0$) for six countries exhibiting shocks. Each point is plotted at the midpoint of the 5-year period it represents. The lines between each data point are intended only to guide the eye. Data: World Population Prospects, 2022 revision \citep{wpp2022methodology}.}
\end{figure}

Short-term shocks to $e_0$ may be a result of numerous causes, such as violent conflicts and economic crises. Figure \ref{fig:life-examples} shows six examples where life expectancy in a country exhibit declines. In Bosnia and Herzegovina, notably, the mortality shock only lasts for one 5-year period, after which life expectancy rebounds to close to the same level as before the shock. In other countries, such as Lebanon and Timor-Leste, decreases in $e_0$ take multiple periods to recover to pre-shock levels. In some countries, lower life expectancy is seen at the beginning or end of the available data. The Republic of Korea had a very low life expectancy in the first time period before exhibiting consistent increases ever since. The Syrian Arab Republic has decreased life expectancy in the final two time periods. These examples illustrate the types of shocks that statistical models of life expectancy need to be able to account for.

\section*{Methods}
\label{section:modeling-framework}

In this methods section, we first introduce a transition model for projecting $e_0$ based on the model of \cite{raftery2013life}. We then extend the model to account for shocks and introduce shrinkage priors to capture shocks. Finally, we discuss interpretation of estimated shock terms and prior settings. To ease exposition we present an abbreviated version of the model set up here, and defer the full specification to the appendix.

To begin, let $\eta_{c,t}$ be defined as the true male period life expectancy at birth in country $c$ and time point $t$. Following \cite{raftery2013life} we assume that the UNPD estimates of $e_0$ are accurate and without error. Statistical modeling of $e_0$ is therefore intended only for generating projections, as modeling is not necessary to reconstruct historical trends. 

\subsection*{Transition model}
In a transition model, the rate of change of an indicator is parameterized by a function incorporating systematic and smoothing components \citep{susmann2022temporal}. The systematic component captures the expected transition from low to high levels of the indicator of interest. The smoothing component captures deceleration or acceleration found in the data that is not explained by the systematic component. For modeling life expectancy $\eta_{c,t}$, a transition model %expressed for the rate of change 
can be written as
\begin{align}
    \label{eq:life-no-shocks}
     \eta_{c,t} = \eta_{c,t - 1} + f(\eta_{c,t - 1}, \bm{\beta}_c) + \epsilon_{c,t},
\end{align}
where $f$ is a function that models the expected rate of change of the indicator as a function of its previous level (this is the systematic component) and $\epsilon_{c,t}$ captures deviations away from the expected rate of change (this is the smoothing component). Existing approaches vary in their choice of transition function $f$. In this paper, we use the B-spline transition function of \cite{susmann2023splines}, which models the rate of change of the indicator as a function of its level using B-splines.  Importantly, $f$ is assumed to be a positive function. That is, $e_0$ is expected to increase until reaching an asymptotic growth regime. Following \cite{raftery2013life}, we assume an asymptotic growth regime in which life expectancy increases linearly at a certain rate. The distortion term $\epsilon_{c,t}$ captures deviations away from this expected trend. Again following \cite{raftery2013life}, we assume that distortions are not autocorrelated and assign a white noise model:
\begin{align}
\label{eq-eps}
    \epsilon_{c,t}|\tau_\epsilon \sim N(0, \tau_\epsilon^2).
\end{align}

\subsection*{Accounting for shocks}
We implement shocks in the transition model by the addition of a shock term. We assume that shocks result in negative changes to the level of the indicator:
\begin{align}
\label{eq:life-shocks}
    \eta_{c,t} = \eta_{c,t - 1} + f(\eta_{c,t - 1}, \bm{\beta}_c) - \delta_{c,t} + \epsilon_{c,t},
\end{align}
where $\delta_{c,t}$ is constrained to be positive. We call this the \textit{shocks model}, where $\delta_{c,t}$ effects the level $\eta_{c,t}$ for one time point.  

We proceed by specifying a prior that most of the $\delta_{c,t}$ terms are near zero, reflecting a prior belief that shocks to the demographic indicators of interest in this paper are rare. We encode this prior belief through the regularized horseshoe prior \citep{piironen2017horseshoe}, a type of Bayesian shrinkage prior. 

To motivate the regularized horseshoe, we first review the definition of the classical horseshoe prior \citep{carvalho2009horseshoe} defined on a parameter $\theta_{c,t}$:
\begin{align}
    \theta_{c,t} \mid  \gamma_{c,t}, \tau &\sim N(0, \tau^2 \gamma_{c,t}^2) \\
    \gamma_{c,t} &\sim C^+(0, 1).
\end{align}
The global scale parameter $\tau > 0$ shrinks all of the $\theta_{c,t}$ parameters toward zero. The local scale parameters $\gamma_{c,t} > 0$ are assigned a heavy-tailed Cauchy prior, allowing individual $\theta_{c,t}$ to escape the global shrinkage. 

The classical horseshoe prior has two drawbacks. First, it is not possible to introduce prior information to regularize large $\theta_{c,t}$. That is, if we have prior information on the expected scale of a non-zero $\theta_{c,t}$ parameter, it is not possible to introduce this within the structure of the horseshoe prior. Second, in practice the complex posterior geometry induced by the horseshoe prior can prevent Markov-chain Monte Carlo (MCMC) algorithms from efficiently sampling from the posterior of models incorporating the classical horseshoe prior \citep{piironen2017horseshoe}.

The regularized horseshoe prior is a related prior with attractive properties that overcome the limitations of the classical horseshoe \citep{piironen2017horseshoe}. The prior is given by
\begin{align}
\theta_{c,t} \mid \gamma_{c,t}, \tau, \vartheta &\sim N(0, \tau^{2} \tilde{\gamma}_{c,t}^2), \\
\tilde{\gamma}_{c,t}^2 &= \frac{\vartheta^2 \gamma_{c,t}^2}{\vartheta^2 + \tau^2 \gamma_{c,t}^2}, \\
\gamma_{c,t} &\sim C^+(0, 1).
\end{align}
where $\vartheta > 0$ is a newly introduced \textit{slab scale} parameter that controls the regularization of large values of $\theta_{c,t}$.  When the $\theta_{c,t}$ are large (that is, they escape the shrinkage imposed by the global shrinkage parameter $\tau$), then the prior on $\theta_{c,t}$ approaches a Gaussian prior with variance $\vartheta^2$. 

The shock terms used for $e_0$ are assigned this regularized horseshoe prior, adapted such that the $\delta_{c,t}$ are constrained to be positive by assuming a half-normal prior:
\begin{align}
\delta_{c,t} \mid \gamma_{c,t}, \tau, \vartheta &\sim N^+(0, \tau^{2} \tilde{\gamma}_{c,t}^2), \\
\tilde{\gamma}_{c,t}^2 &= \frac{\vartheta^2 \gamma_{c,t}^2}{\vartheta^2 + \tau^2 \gamma_{c,t}^2}, \\
\gamma_{c,t} &\sim C^+(0, 1).
\end{align}
The positivity constraint is because while we expect short-term and reversible decreases in life expectancy, we do not expect similarly short-term increases in life expectancy.

We propose analyzing the magnitude of the estimated shocks relative to the size of ``regular'' fluctuations in life expectancy, as quantified by the deviation term $\epsilon$. In particular, we define a ``shock" as when the value of $\delta$ is greater than twice the marginal standard deviation of $\epsilon$: $\delta > 2 \mathrm{sd}(\epsilon) := \delta^*$. In this case study, given the expression for $\epsilon$ in Equation \eqref{eq-eps}, we define the \textit{shock threshold} $\delta^* = 2\cdot \text{sd}(\epsilon) = 2\cdot \hat{\tau}_\epsilon$, with  $\hat{\tau}_\epsilon$ the posterior median estimate of ${\tau}_\epsilon$. 

To complete the model specification we must choose prior distributions for $\tau$ and $\vartheta$. Details are provided in the appendix. In summary, we follow \cite{piironen2017horseshoe} in their choice of prior family for each parameter and propose a principled way to set the prior hyperparameters. First, we calibrate the degree of prior regularization for any large values of $\delta_{c,t}$ that escape the horseshoe by assuming that the magnitude of a shock cannot exceed 100 life-expectancy years and setting a target probability of $10\%$ of a shock exceeding 20 life-expectancy years. Second, we tune the remaining prior hyperparameters based on the probability of the $\delta_{c,t}$ being large, relative to shock threshold $\delta^*$. Specifically, we assume that  $P(\delta_{c,t} > \delta^*) \approx 0.5\%$. We do a sensitivity analysis and find that our substantive results are not sensitive to prior choice. 

\subsection*{Computation} The model was implemented in the \texttt{Stan} programming language \citep{stan2023, cmdstanr2022}, and all analyses were conducted using the \texttt{R} statistical computing environment \citep{r2022}. Samples were drawn from the joint posterior distribution of the model parameters using the Hamiltonian Markov-Chain Monte Carlo (MCMC) algorithm with the No-U-Turn sampler \citep{hoffman2014no}. For all results, $4$ MCMC chains were run with $500$ initial samples in the warm-up stage and $1000$ samples in the sampling stage. Convergence of the MCMC results was assessed using the split $\hat{R}$ diagnostic introduced in \cite{vehtari2021rhat}. 

\section*{Results}
\begin{figure}[h]
    \includegraphics[width=\columnwidth]{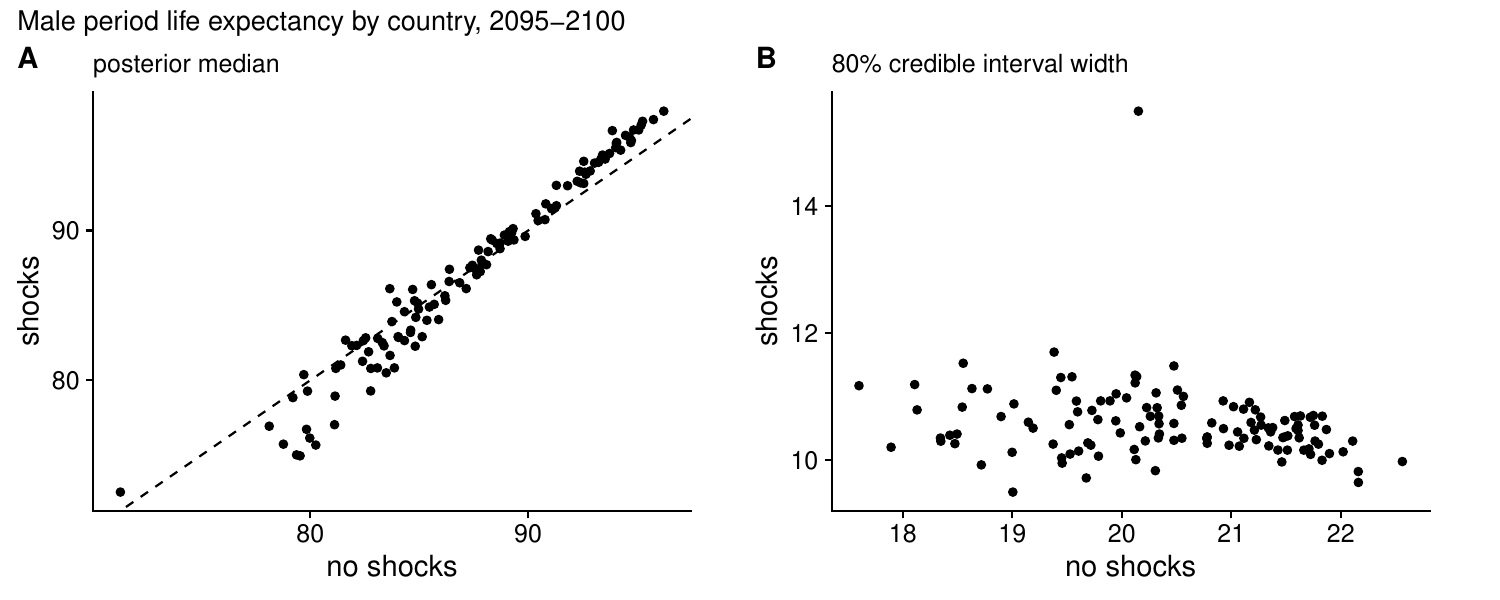}
    \caption{\label{fig:life-projection-comparison} Posterior medians (A) and 80\% projection interval widths (B) for male period life expectancy at birth by country in 2095-2100 for the model with and without shocks included.}
\end{figure}

The models with and without shock were fitted to UNPD $e_0$ estimates up to 2015-2020 and used to produce projections until 2095-2100. Resulting projections for all countries are included in the appendix. Projections for selected countries are shown in Figure~\ref{fig:life-fit-examples}; projections for all countries are included in the appendix. Cuba, Tajikistan, Estonia, and Jordan are examples of countries where the point estimates for $e_0$ are similar between the two models, while Yemen, Republic of Moldova, Syrian Arab Republic, and Papua New Guinea are examples of countries with differences. For these countries, the model that includes shocks has narrower projection intervals than the model without shocks. For a wider comparison, Figure~\ref{fig:life-projection-comparison} compares point estimates and width of projection intervals for the period 2095-2100 for all countries. In long-term projections, including shocks in the model resulted in narrower projection intervals for all countries. Point projections are similar though slightly lower (higher) for low (high) $e_0$ countries in the model with shocks. 

\begin{figure}[h]
    \includegraphics[width=\columnwidth]{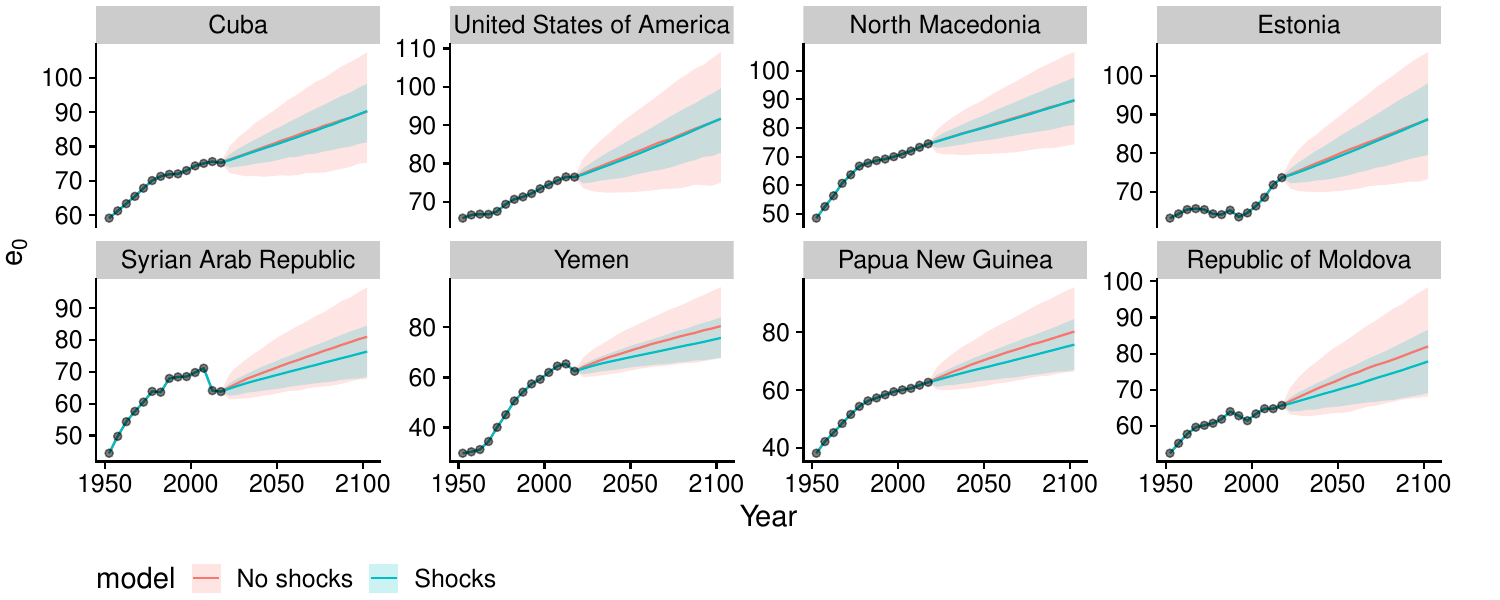}
    \caption{\label{fig:life-fit-examples} Projections of $e_0$ from the model without shocks \eqref{eq:life-no-shocks} and with shocks \eqref{eq:life-shocks}, for countries with the smallest (top row) and largest (bottom row) differences in posterior median projected $e_0$ in 2095-2100. Estimates (dots), projections (line), and 80\% projection intervals (shaded areas) are shown. }
\end{figure}

\begin{figure}[h]
    \centering
    \includegraphics[width=\columnwidth]{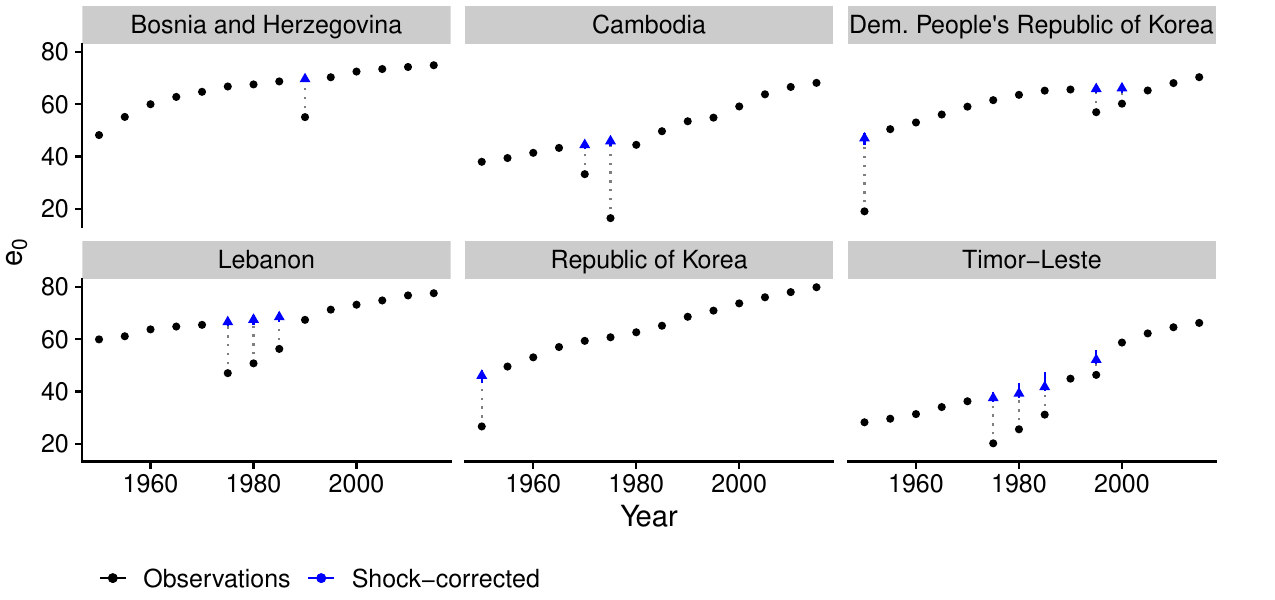}
    \caption{Six countries with the largest estimated detected shocks. Shocks are illustrated by plotting ``shock-corrected'' estimates, given by the observed $e_0$ minus the shock $\delta_{c,t}$, with error bars showing the estimated 95\% credible interval.}
    \label{fig:largest-shocks}
\end{figure}

The inclusion of the shock terms allows for data-driven discovery of country-years that likely exhibit large life-expectency fluctuations. The posterior median of $\tau_\epsilon$ was $0.82$ years, suggesting a threshold for shocks of $\delta^* = 2 \cdot \mathrm{sd}(\epsilon) = 1.64$ years. A total of 38 country-years were identified with posterior probability $P(\delta_{c,t} > \delta^*) > 97.5\%$, reflecting high posterior probability of there being a shock. These country-years comprise 22 unique countries. The six countries with the largest estimated shock (in terms of the posterior median of $\delta_{c,t}$) are shown in Figure \ref{fig:largest-shocks}. The periods identified in the six countries (Bosnia and Herzegovina, Cambodia, Dem. People's Republic of Korea, Lebanon, Republic of Korea, and Timor-Leste) have documented crises. A similar figure for all countries that had shocks detected is included in the appendix (Figure \ref{fig:all_largest_shocks}).

To investigate model performance, out-of-sample model validations were used to compare the predictive performance of the life expectancy model with and without shocks. All observations after 2005-2010 were held out, and each model fitted to the remaining observations. The fitted model was then used to predict the held-out observation from 2015-2020. Thus, the validations reflect 10-year predictive accuracy. The projections were evaluated in terms of empirical coverage of the 80\% projection intervals, median 80\% projection interval width, median error (ME), and median absolute error (MAE).

Results from the out-of-sample 10-year prediction validations are shown in Table~\ref{tab:life-validations} by world region. The models with shocks had uniformly smaller 80\% projection interval widths than the model without shocks, while having comparable median errors and median absolute errors. The 80\% projection intervals of the model without shocks were conservative, with empirical coverage of 96.7\% of the held-out data. The projection intervals of the shocks model were nearly optimal at $75.2\%$ in the 10-year ahead forecasting exercise. Coverage for the model with shocks is reasonable for all regions but we note some asymmetry in the percentages above and below the projection intervals, in particular for Europe, where results suggests that projections from 2005-2010 are biased. This is due to limitations of the transition model used (see discussion).

\begin{table}[h]
    \centering
    \begin{tabular}{|lllrrrrrr|}
        \hline
        \multicolumn{5}{|c}{} & \multicolumn{4}{c|}{80\% projection Interval}  \\
        Model & Region & $n$ & ME & MAE & \% Below & \% Included & \% Above & PI Width \\
        \hline
        shocks & Africa & 11 & 0.49 & 1.14 & 9.1\% & 72.7\% & 18.2\% & 3.12\\
               & Americas & 25 & -0.07 & 0.67 & 12.0\% & 88.0\% & 0.0\% & 3.12\\
               & Asia \& Oceania & 50 & 0.62 & 0.86 & 4.0\% & 72.0\% & 24.0\% & 3.13\\
               & Europe & 35 & 1.10 & 1.10 & 0.0\% & 71.4\% & 28.6\% & 3.08 \\
         & \textit{Overall} & \textit{121} & \textit{0.57} & \textit{0.92} & \textit{5.0\%} & \textit{75.2\%} & \textit{19.8\%} & \textit{3.11} \\ 
        no shocks & Africa & 11 & 0.42 & 1.27 & 0.0\% & 100.0\% & 0.0\% & 7.55\\
                  & Americas & 25 & -0.22 & 0.68 & 0.0\% & 100.0\% & 0.0\% & 7.61\\
                  & Asia \& Oceania & 50 & 0.67 & 1.02 & 4.0\% & 92.0\% & 4.0\% & 7.56\\
                  & Europe & 35 & 0.76 & 0.78 & 0.0\% & 100.0\% & 0.0\% & 7.64\\
         & \textit{Overall} & \textit{121} & \textit{0.61} & \textit{0.86} & \textit{1.7\%} & \textit{96.7\%} & \textit{1.7\%} & \textit{7.60} \\ 
        \hline
    \end{tabular}
    \caption{\label{tab:life-validations}Validation results for the male period life expectancy at birth model, with shocks \eqref{eq:life-shocks} and without shocks \eqref{eq:life-no-shocks}. Included validation metrics are median error (ME), and median absolute error (MAE), the \% of observations below, above, and included within the 80\% projection interval, and the mean width of the 80\% projection interval.}
\end{table}

\section*{Discussion}
\label{section:discussion}
Many demographic and health indicators smoothly evolve over time apart from rare large deviations, which we refer to as shocks. Statistical models that assume smooth trends may produce biased or poorly calibrated projections when confronted with data that include shocks. We proposed a general modeling approach for handling shocks based on the use of Bayesian shrinkage priors that explicitly accounts for shocks without the need for potential crisis periods to be identified a-priori. In a case study of male life expectancy at birth, the benefits of the approach were clear: adding shocks yielded shorter projection intervals without sacrificing good empirical coverage in validation exercises. 

In this paper we focused on projections with shock terms that account for the possibility of shocks as observed globally since 1950. If of interest, the model may also be used to produce shock-free projections, where the shock term is removed from the projections for all or a subset of countries. Vice versa, the model can be used to produce updated projections for countries with ongoing crises, using a posterior predictive density for the local parameter of the shock term, to quantify the shock expected in crisis settings. Such an approach could improve upon the current practice of having to manually update prior settings of transition model parameters \citep{wpp2022methodology}. In addition, the projections in this paper are informed by $e_0$ estimates up to the period 2015-2020 only, and therefore do not incorporate excess mortality associated with the Covid-19 pandemic. If estimates are updated to included country-periods where $e_0$ is affected by Covid-19, the model with shocks can be used to capture the associated decline in $e_0$. 

Model validation suggests that the model shocks improves upon the model without shocks in terms of reducing the width of projection intervals. An issue identified in these exercises is a bias and resulting lower-than-nominal coverage for projections starting in 2005 for some countries, where the models with and without shock underpredict $e_0$ gains. We believe that these results can be improved by reconsidering the transition model. In particular, the countries under consideration experience prolonged periods where $e_0$ does not increase. Such stalls could be captured by autocorrelated deviation terms. 

While in this work our approach for shocks was illustrated by capturing temporary decreases in life expectancy, the overall approach using shrinkage priors is applicable more generally. We expect that the addition of shock terms modeled by shrinkage priors may have advantages in the greater class of models commonly used for producing estimates and projections of demographic and health indicators, referred to as the Temporal Models for Multiple Populations (TMMPs) model class \citep{susmann2022temporal}. For example, models that include covariates, such as GBD's approach to projecting $e_0$, can consider the addition of a shock term on the input covariates and/or the outcome \citep{foreman2018}. 

\subsection*{Acknowledgements}
We would like to thank Patrick Gerland, Mark Wheldon, Vladimira Kantorova, Kristin Bietsch, John Stover, and Emily Sonneveldt for helpful comments. 
This work was supported, in whole or in part, by the Bill \& Melinda Gates Foundation (INV-00844). Under the grant conditions of the Foundation, a Creative Commons Attribution 4.0 Generic License has already been assigned to the Author Accepted Manuscript version that might arise from this submission.

\bibliography{bibliography}

\clearpage 

\section*{Appendix}
\subsection*{Full model specification}
We write the process model for $e_0$ using the Temporal Models for Multiple Populations (TMMPs) framework \citep{susmann2022temporal}. The full model  includes systematic, shock, and stochastic smoothing components:
\begin{align}\label{appendix:life-full-spec}
\eta_{c,t} = \eta_{c,t - 1} + f(\eta_{c,t - 1}, \bm{\beta}_c) - \delta_{c,t} + \epsilon_{c,t}.
\end{align}
For the function $f$, referred to as the \textit{transition function}, we adopt the B-spline transition function of \cite{susmann2023splines}, which models the rate of change of the indicator as a function of its level using flexible B-splines. 
The transition function is given by, with $\bm{\beta}_c = \bm{\lambda}_c, \bm{\beta}^\prime_c$, 
\begin{align}
    f(\eta, \bm{\beta}_c) = \sum_{k=1}^K \beta_{c,k}^\prime B_k((\eta - \lambda_{\ell}) / (\lambda_u - \lambda_{\ell})),
\end{align}
which is a weighted combination of $K = 6$ B-spline functions $B_k$, evaluated at $\eta$ rescaled using $\lambda_\ell = 15$ and $\lambda_u = 110$. The spline knots are evenly spaced between $15$ and the maximum observed life expectancy in the dataset. We impose constraints on the spline coefficients $\beta_{c,k}^\prime$, reproducing constraints introduced by \cite{raftery2013life}. Specifically, the first $K - 3$ spline coefficients are constrained to be between $0$ and $10$. This follows \cite{raftery2013life}, who assumed the maximum rate of change of $e_0$ to be less than 10. The final $3$ coefficients are set to be equal and constrained to be less than $1.15$, again following \cite{raftery2013life} who set the same upper limit for the asymptotic rate of increase in $e_0$. These constraints for the spline coefficients ensure a positive rate of change from $15$ to $110$, with the rate of change constrained to be less than 1.15 once $e_0$ reaches 110 years. Formally, these constraints were implemented as
\begin{align}
    \beta^\prime_{c,i} &= 0.001 + (10 - 0.001) \times \mathrm{logit}^{-1}(\beta^*_{c,i}), \text{ for } i = 1, \dots, K - 3, \\
    \beta^\prime_{c,i} &= 0.001 + (1.15 - 0.001) \times \mathrm{logit}^{-1}(\beta^*_{c,K - 2}), \text { for } i = K - 2, \dots, K.
\end{align}
Hierarchical models were then placed on the unconstrained parameters $\beta_{c,i}^*$:
\begin{align}
    \beta_{c,i}^* \sim N(\beta_{i}^*, \sigma_{i}^2), \text{ for } i = 1, \dots, K - 2.
\end{align}
The priors on $\beta_{i}^*, \sigma_{i}^2$ were
\begin{align}
    \beta_i^* &\sim N(0, 15), \\
    \sigma_i &\sim N^+(0,5). 
\end{align}
The stochastic smoothing term $\epsilon_{c,t}$ is assigned a white-noise prior for all time points:
\begin{align}
    \epsilon_{c,t} &\sim N(0, \tau_{\epsilon}^2), \\
    \tau_{\epsilon}^2 &\sim \mathrm{InvGamma}(0.1, 0.1).
\end{align}
The shock terms use for $e_0$ are assigned a regularized horseshoe prior, with the $\delta_{c,t}$'s constrained to be positive:
\begin{align}
\delta_{c,t} \mid \gamma_{c,t}, \tau, \vartheta &\sim N^+(0, \tau^{2} \tilde{\gamma}_{c,t}^2), \\
\tilde{\gamma}_{c,t}^2 &= \frac{\vartheta^2 \gamma_{c,t}^2}{\vartheta^2 + \tau^2 \gamma_{c,t}^2}, \\
\gamma_{c,t} &\sim C^+(0, 1).
\end{align}

\paragraph{Specification of the shock term and shock prior tuning parameters}\label{appendix:hsprior}

To complete the model specification we must choose prior distributions for $\tau$ and $\vartheta$. We follow \cite{piironen2017horseshoe} in considering priors for $\tau$ of the form
\begin{align}
    \tau \sim C^+(0, \tau_0^2),
\end{align}
for a fixed choice of $\tau_0$, and for the slab scale parameter $\vartheta$, we consider priors of the form
\begin{align}
    \vartheta^2 \sim \text{Inv-Gamma}(\nu/2, \nu s^2 / 2).
\end{align} 
Taken together, we need to choose values for the hyperparameters $\tau_0$, $\nu$, and $s$. \cite{piironen2017horseshoe} includes a principled method for choosing $\tau_0$ in the context of generalized linear models based on the expected number of non-zero coefficients. For our proposed model class, such guidance is not applicable. As a principled way to set the hyperparameters, we propose to choose suitable values for the regularized horseshoe prior tuning parameters in two steps. First, we calibrate the degree of prior regularization for any large values of $\delta_{c,t}$ that escape the horseshoe. Second, we tune the remaining prior hyperparameters based on the probability of the $\delta_{c,t}$ being large, relative to shock threshold $\delta^*$. 

Reasonable constraints exist to guide the choice of a prior on large values of $\delta_{c,t}$. Let  $\delta^{(\text{large})}$ denote a typical $\delta_{c,t}$ that has escaped global shrinkage to capture a shock; more formally, $\delta_{c,t} \xrightarrow[]{\gamma_{c,t} \to \infty} \delta^{(\text{large})}$. When following \cite{piironen2017horseshoe} and considering priors on the slab scale parameter $\vartheta$ of the form
\begin{align}
    \vartheta^2 \sim \text{Inv-Gamma}(\nu / 2, \nu s^2 / 2),
\end{align}
then the prior on  $\delta^{(\text{large})}$ is given by a half-Student-$t$ distribution:
\begin{align}
  \delta^{(\text{large})} \sim  \text{Student-}t^+_{\nu}(0, s^2). 
%        \delta_{c,t} \xrightarrow[]{\gamma_{c,t} \to \infty} \text{Student-}t^+_{\nu}(0, s^2). 
\end{align}
To guide our choice of $\nu$ and $s$, we first note that is quite reasonable to expect that life expectancy can never change by more than 100 years between two five-year periods. By consequence, the shocks can not exceed $100$ years. The prior should thus be set such that $P( \delta^{(\text{large})} > 100) \approx 0$. To further constrain the prior, we set a target probability of $10\%$ of the shocks exceeding 20 years: $P( \delta^{(\text{large})} > 20) \approx 0.1$. This encodes a prior belief that most shocks will have a magnitude of less than 20 years, but that there should be significant prior mass on the possibility of even larger shocks. This relationship suggests setting $\nu = 6$ and $s^2 = 10$ as reasonable prior hyperparameter values.

\begin{figure}
    \includegraphics[width=0.9\textwidth]{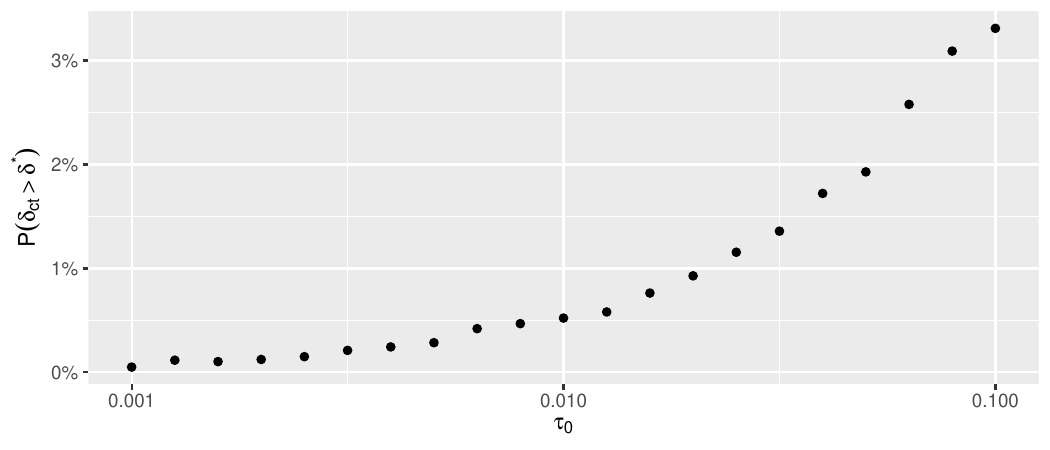}
    \caption{\label{fig:prior_predictive} Estimated $P(\delta_{c,t} > \delta^*)$ based on various prior choices of global scale parameter $\tau_0$. 
    }
\end{figure}

To find a reasonable setting for the global scale parameter $\tau_0$, that controls the probability of the shock terms being large, we turn to prior predictive simulations. Recall that a shock is defined as when $\delta_{c,t} > 2 \mathrm{sd}(\epsilon) := \delta^*$. As an initial estimate of the standard deviation of the smoothing term, we fit the model without a shock term and calculated $\mathrm{sd}(\epsilon)$ to be the posterior median of $\tau_\epsilon$. 
%This yields a working value for the shock threshold of $\tilde{\delta}^* = 3.56$ years. 
Note that sets a high, conservative threshold for what constitutes a shock, because the standard deviation of $\epsilon$ will be higher in a model that does not include a separate shock term. Prior predictive simulations can be used to estimate $P(\delta_{c,t} > \delta^*)$ conditional on various choices of the global scale hyperparameter $\tau_0$ and using previously set values for $\nu$ and $s^2$. The results are shown in Figure \ref{fig:prior_predictive}, and indicate that varying $\tau_0$ from $0.001$ to $0.1$ increases $P(\delta_{c,t} > \delta^*)$ from approximately $0.05\%$ to $3.3\%$. The choice $\tau = 0.01$, corresponding to $P(\delta_{c,t} > \delta^*) \approx 0.5\%$, appears reasonable. 

\begin{figure}
    \centering\includegraphics{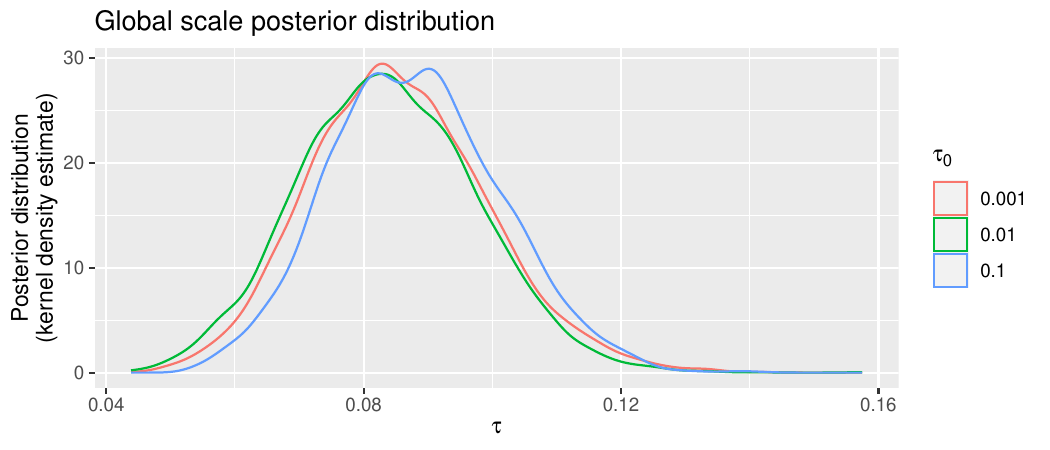}
    \caption{Posterior distributions of the global scale parameter $\tau$ for choices of hyperparameter $\tau_0 \in \{ 0.001, 0.01, 0.1 \}$.}
    \label{fig:global_scale_posterior}
\end{figure}

\paragraph{Sensitivity analysis} We fitted the full model with values $\tau_0 \in \{ 0.001, 0.01, 0.1 \}$ to check for sensitivity to the prior choice. 
Posterior distributions of the global scale parameter $\tau$ for choices of hyperparameter $\tau_0 \in \{ 0.001, 0.01, 0.1 \}$ are shown in Figure \ref{fig:global_scale_posterior}. The posterior is similar for all choices of $\tau_0$.

\clearpage 

\subsection*{Supplementary figures}

\begin{figure}[h]
    \centering
    \includegraphics[width=\textwidth]{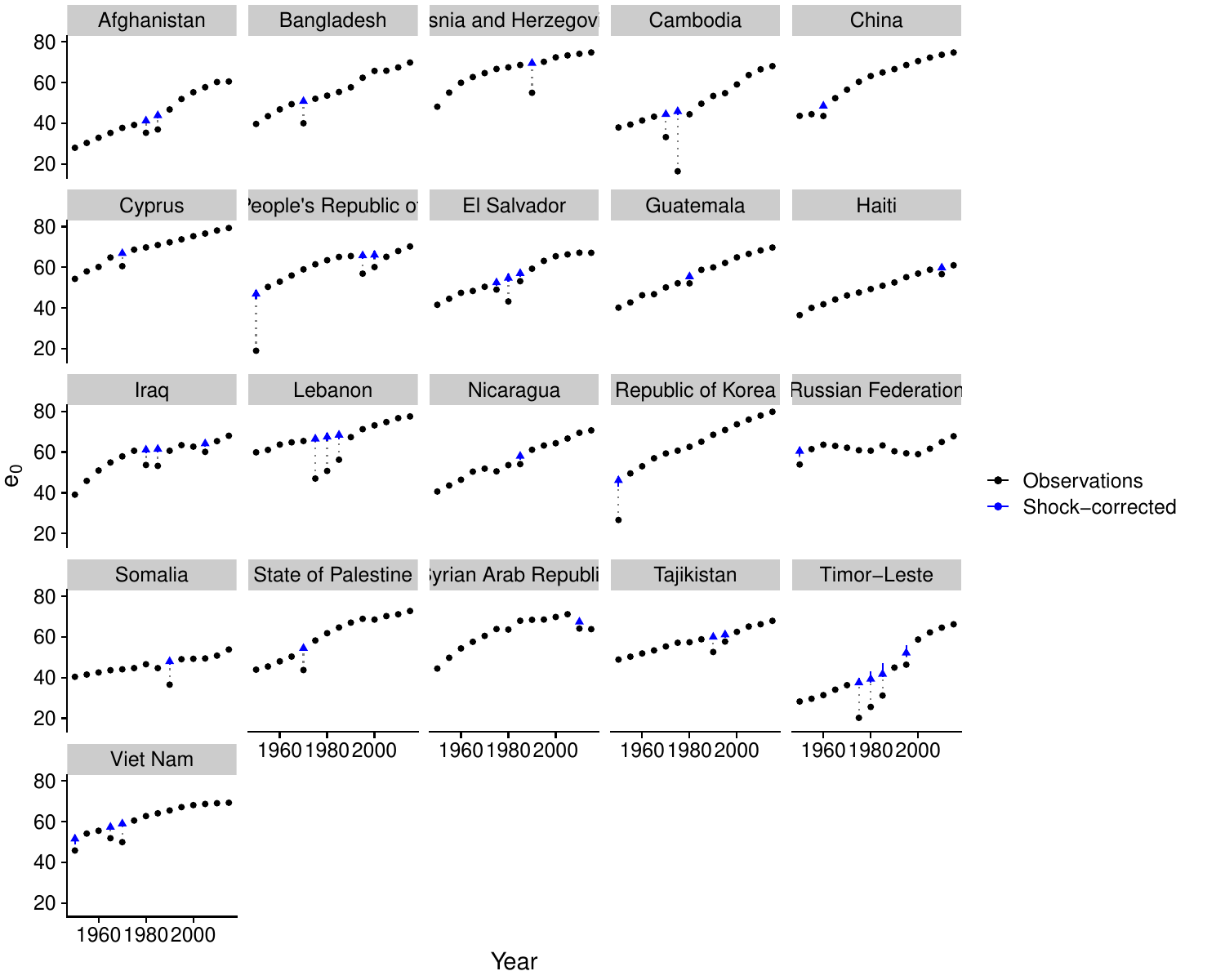}
    \caption{All countries with detected shocks. Shocks are illustrated by plotting ``shock-corrected'' estimates, given by the observed $e_0$ minus the shock $\delta_{c,t}$, with error bars showing the estimated 95\% credible interval.}
    \label{fig:all_largest_shocks}
\end{figure}

\clearpage

\begin{figure}[h]   \centering
    \caption{ Projections of $e_0$ for all countries from the model without shocks \eqref{eq:life-no-shocks} and with shocks \eqref{eq:life-shocks}. Estimates (dots), projections (line), and 80\% projection intervals (shaded areas) are shown. }
\end{figure}

\clearpage

\includepdf[pages=1-,nup=1x4,pagecommand=,width=\columnwidth]{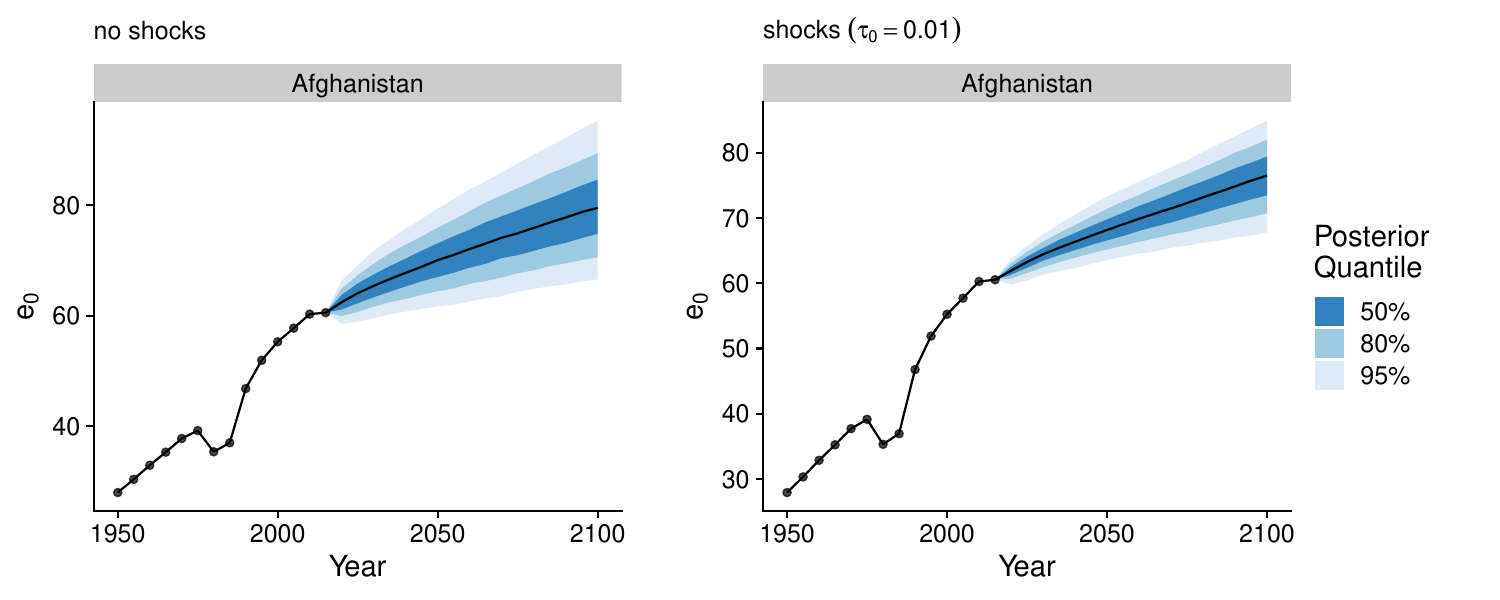}

\begin{figure}[h]   \centering
    \caption{ Projections of $e_0$ for all countries from the validation exercise using the model without shocks \eqref{eq:life-no-shocks} and with shocks \eqref{eq:life-shocks}. Estimates (dots), projections (line), and 80\% projection intervals (dotted lines) are shown. }
\end{figure}

\clearpage

\includepdf[pages=1-,nup=2x4,pagecommand=,width=0.4\columnwidth]{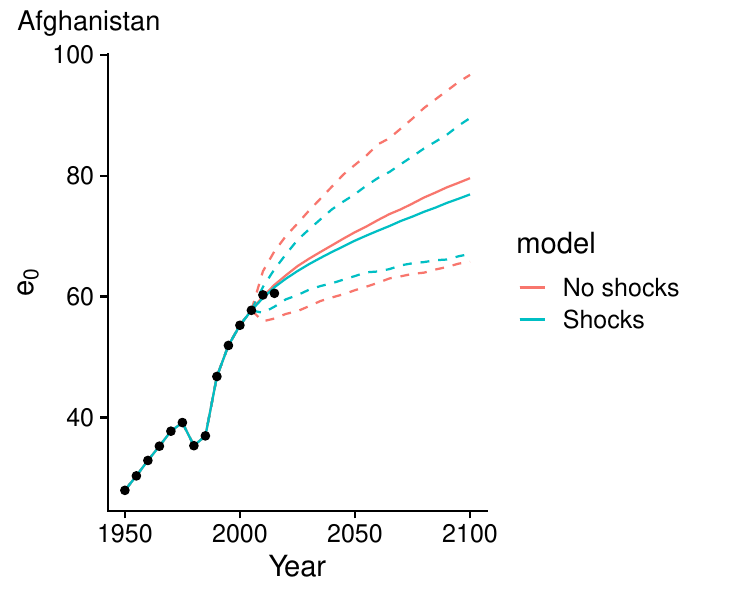}

\end{document}